\documentclass[]{acm_proc_article-sp}
\usepackage{url}

\usepackage{bigpage}
\usepackage{bcprules}
\usepackage{mathpartir}
\usepackage{listings}
\usepackage{mathtools}
\usepackage{amsfonts}
\usepackage{cmll}
\usepackage{latexsym}
\usepackage{amssymb}
\usepackage{caption}

\newcommand{\id}[1]{\texttt{#1}}

\newcommand{\bc}{\mathbin{\mathbf{::=}}}
\newcommand{\bm}{\mathbin{\mathbf\mid}}

\newlength{\ltext}
\newlength{\lmath}
\newlength{\cmath}
\newlength{\rmath}
\newlength{\rtext}

\newenvironment{grammar}{
  \[
  \begin{array}{l@{\quad}rcl@{\quad}l}
  \hspace{\ltext} & \hspace{\lmath} & \hspace{\cmath} & \hspace{\rmath} & \hspace{\rtext} \\
}{
  \end{array}\]
}

 \numberwithin{equation}{subsection}

\newcommand{\papertitle}{Linear types can change the blockchain!}

\title{\papertitle}

\author{
L.G. Meredith\\
  \affaddr{CSO, Synereo}\\
  \email{\fontsize{8}{8}\selectfont greg@synereo.com}
}

\begin{document}
\lstset{language=}

\setlength{\topmargin}{0in}
\setlength{\textheight}{8.5in}
\setlength{\parskip}{6pt}


\begin{abstract}
\normalsize{ 

  We present an interpretation of classical linear logic in terms of operations on the blockchain.

}

\end{abstract}


\maketitle




\section{Background and Motivation}

Anyone who understands the current economic, sociological, and
technological situation is likely to be very excited by what the
blockchain technology promises. Anyone who has actually had to work
with the blockchain in real situations with mission-critical exchanges
on the line is very likely to be motivated to find a more scalable and
reliable architecture for the blockchain. This paper takes a few key
steps towards finding a way to explain and test a hypothesis that
linear proofs provide the basis for a much more scalable architecture
for the blockchain. For background on what is meant here by linear
proofs, \cite{DBLP:conf/lics/AbramskyM99} interprets them in terms of
games, while \cite{DBLP:journals/tcs/Abramsky93} interprets them in
terms of traditional computational calculi like the lambda calculus.

A linear proof is a formal structure representing a proof of a formula
in linear logic \cite{DBLP:journals/tcs/Girard87}.  The Curry-Howard
isomorphism \cite{Sorensen98lectureson} tells us that formulae are
types (as in data types in a programming language), and that proofs
are programs. This is a very broad and deep idea. In the 90's, for
example, Abramsky extended it to proofs as processes
\cite{DBLP:journals/tcs/Abramsky94}, which Wadler was only very
recently able to realize as a correspondence between linear proofs and
{$\pi$}-calculus processes \cite{DBLP:journals/jfp/Wadler14}. In this
context it means that linear proofs provide a representation of both
data (blocks) and program (executable transactions) that gives several
advantages over the current choices made by the blockchain.

The blockchain is a great example of data that is also program; it's a
giant ledger spread out over the Internet, that's made of a bunch of
distributed, but interacting servers \cite{Nakamoto2008}. To become
more scalable, and reliable, both ledger and servers will need certain
characteristics of data/program that have to do with a property called
compositionality. Scalability is always all about being able to build
composite systems from components. For example, if we can prove that
sections of the blockchain can be safely isolated from other sections,
for example, if all blocks necessary to prove that Alice has
sufficient funds to send M btc to Betty, can be isolated from the
blocks necessary to prove that Alfred has sufficient funds to send N
btc to Bob, then Alice and Betty, and Alfred and Bob can safely work
with projections of the blockchain, and thus complete their
transactions, not only in isolation of each other, but without the
onerous need to sync the entire blockchain.

One analogy is the use of separation logic (a child of linear logic)
\cite{DBLP:conf/vstte/Reynolds05} to prove things about the structure
of the heap which can, in turn, be used to guarantee that two threads
can operate at the same time safely. The blockchain is like the
heap. The Alice - Betty and Alfred - Bob transactions are like the two
threads. A proof that the heap is of the form $H_1 \otimes H_2$
together with a proof that $T_1 : H_1 \rightarrow H_1'$, $T_2 : H_2
\rightarrow H_2'$ constitutes a proof that $T_1 \otimes T_2$ (thread
$T_1$ running concurrently with $T_2$) operate effectively in
isolation and thus safely. Likewise, a proof that the blockchain is of
the form $B_1 \otimes B_2$ (none of the transactions in $B_1$ connect to
addresses in $B_2$ and vice versa), together with a proof that
$\mathsf{AliceBetty} : B_1 \rightarrow B_1'$ (this txn uses only addresses
in $B_1$), and $\mathsf{AlfredBob} : B_2 \rightarrow B_2'$ (this txn uses
only addresses in $B_2$) constitutes a proof that $\mathsf{AliceBetty}
\otimes \mathsf{AlfredBob}$ can operate with isolated projections of the
blockchain.

If the blockchain is built using the primitives of linear logic, it
becomes easier and easier to construct these proofs, but also to
construct the blockchain, itself, in terms of smaller blockchains.

\section{Interpreting Linear Proofs as Operations on the Blockchain}
Here's the most basic interpretation.

\begin{mathpar}
  \inferrule* {} {\vdash 1blkchnaddr : \underbrace{A \otimes \ldots \otimes A}_{M}} 
\end{mathpar}

is a statement that there are $M$ $A$'s available at the address,
$1blkchnaddr$. $A$'s can be any resource, BTC's, AMP's, DogeCoin, etc.

\begin{mathpar}
  \inferrule* {} {\vdash txn : \underbrace{A \otimes \ldots \otimes A}_{M}  \multimap \underbrace{B \otimes \ldots \otimes B}_{N} } 
\end{mathpar}

is a statement that $txn$ will generate $N$ $B$'s if provided $M$ $A$'s.

Terminologically, we say that $1blkchnaddr$ is a \emph{witness} or a
\emph{proof} of $A \otimes \ldots \otimes A$, and similarly, that $txn$
is a witness or proof of $A \otimes \ldots \otimes A \multimap B \otimes \ldots \otimes B$. Given two such
proofs we can use the cut rule of linear logic to produce a proof

\begin{mathpar}
  \inferrule* {} {\vdash txn( 1blkchnaddr ) : B \otimes \ldots \otimes B}
\end{mathpar}

where $txn( 1blkchnaddr )$ is a new address in the blockchain
constructed from the information in txn together with
$1blkchnaddr$. This should look remarkably like function application,
because it is.

Notice that now we can see, recursively, what a proof of a statement
like $\vdash 1blkchnaddr : A \otimes \ldots \otimes A$ looks like. In
most cases it will be a proof made from a previous application of the
cut rule.  This tree of cuts will trace all the way back to some
genesis block -- which is the only other way to have a proof of a
statement like $\vdash 1blkchnaddr : A \otimes \ldots \otimes A$.

Now, where does the return address associated with $txn( 1blkchnaddr
)$ come from? To see this we have to look into the mechanics of
$\multimap$ (called linear implication or, more affectionately,
lollipop).

\begin{equation*}
  A \multimap B = A^{\perp} \parr B
\end{equation*}

Linear implication is decomposed much like classical implication in
terms of a negation ($A^{\perp}$) and a disjunctive connective
($A \parr B$). It is literally an expression capturing the sentiment
we need $A$ to get $B$, or $B$ comes with a cost of $A$. The use of a
proof rule for these kinds of links looks like

\begin{mathpar}
  \inferrule* {\vdash \Gamma, 1blkchnaddr : A^{\perp}, 2blkchnaddr : B}{\vdash \Gamma, 1blkchnaddr \parr 2blkchnaddr : A^{\perp} \parr B}
\end{mathpar}

where we snuck in the rest of the blockchain as $\Gamma$. As we saw above, $A^{\perp} \parr B = A \multimap B$; so, we can write

\begin{mathpar}
  \inferrule* {\vdash \Gamma, 1blkchnaddr : A^{\perp}, 2blkchnaddr : B}{\vdash \Gamma, 1blkchnaddr \multimap 2blkchnaddr : A \multimap B}
\end{mathpar}

Now we see that forming a $txn$ comes with the requirement to provide
an address where $A$'s will be sent and an address where $B$'s will be
received. To complete the picture, applying the cut rule will create
the txn that links an address, say $3blkchnaddr$ with an $A$ in it, to
$1blkchnaddr$, resulting in $\vdash \Gamma, 2blkchnaddr : B$. Expanding on these
intuitions, we can see that the rules of classical linear logic
correspond exactly to a specification of operations on the blockchain.

\subsection{Linear Sequents}

In more detail, a proof rule in linear logic is usually written in
terms of a transformation,

\begin{mathpar}
  \inferrule* {S_1, S_2, \ldots, S_N} {S}
\end{mathpar}

taking \emph{sequent}s $S_1, \ldots, S_N$ to a \emph{sequent} $S$, where
a sequent is of the form

\begin{mathpar}
  \inferrule* {} {\vdash \Gamma, t_1 : A_1, \ldots, t_N : A_N}
\end{mathpar}

A sequent is really just a statement about what is distributed where in an instance of the blockchain.

\begin{itemize}
  \item $t_1, \ldots, t_N$ are either addresses or programs that take addresses as input; they constitute the "focus" of the proof rule, or where the action is going to happen.
  \item $A_1, \ldots, A_N$ are (types built from) the different types of coin
  \item $\Gamma$ is the rest of the blockchain -- it is necessary to establish the distribution of resources we see at $t_1, \ldots, t_N$, but it's not the focus of the operation of the proof rule. 
\end{itemize}

Putting it all together, a proof rule of the form

\begin{mathpar}
  \inferrule* {S_1} {S_2}
\end{mathpar}

is then a statement about how the blockchain in state $S_1$ goes to a
blockchain in state $S_2$. If you think about it, that's just what we
need to reason about transactions. In a transaction where Alice sends
Betty $N$ coin, we can think of the transaction as a rule that takes a
blockchain in a state where Alice has $N$ btc to a blockchain in a
state where Betty has $N$ btc.

\subsection{The Multiplicatives}
Linear logic, however, allows to build bigger blockchains from smaller
ones, and manages the dependency and information flow so that
everything remains consistent. Here's an example. The proof rule for
the tensor $A \otimes B$ looks like this

\begin{mathpar}
  \inferrule* {\vdash \Gamma, t : A,  \vdash \Delta, u : B}{\vdash \Gamma, \Delta, t \otimes u : A \otimes B}
\end{mathpar}

It says that if you have one blockchain, $\vdash \Gamma, t : A$, and
another completely independent blockchain, with a totally separate
address space, $\vdash \Delta, u : B$, then you can make a new one

\begin{mathpar}
  \inferrule* {} {\vdash \Gamma, \Delta, t \otimes u : A \otimes B}
\end{mathpar}

in which you just combine all the data of assignments of addresses to
resources in $G$ and $H$ in one big blockchain, $G,H$, and you can make a
kind of composite address (or program), $t \otimes u$, at which can be found
the combined $A \otimes B$ resource.

Now, comparison of the par ( $A \parr B$ ) rule, which establishes transaction links, is even more illuminating.

\begin{mathpar}
  \inferrule* {\vdash \Gamma, t : A, u : B}{\vdash \Gamma, t \parr u : A \parr B }
\end{mathpar}

This rule insists that the transaction link,  $t \parr u$, is made in the same piece of the blockchain, $\Gamma$.

The piece of the puzzle that interprets commitment to and execution of
transactions is the cut rule. If $1blkchnaddr \multimap 2blkchnaddr$
is a transaction waiting to happen, so to speak, $txn( 3blkchnaddr,
1blkchnaddr -o 2blkchnaddr )$ is the commitment to carry out the txn
against the blockchain. Likewise, cut-elimination, also called
proof-normalization, which corresponds to computation, via
Curry-Howard, constitutes the execution of the transaction on the
blockchain that results in the assignment $\vdash \Gamma, 2blkchnaddr
: B$ after execution. Someone familiar with functional programming
might interpret

$txn( 3blkchnaddr, 1blkchnaddr \multimap 2blkchnaddr )$

as

$apply( 1blkchnaddr \multimap 2blkchnaddr, 3blkchnaddr )$

making the correspondence to function application, and the
correspondence between proof normalization and {$\beta$}-reduction
explicit.

The fragment of linear logic that includes, $A^{\perp}$, $A \otimes
B$, $A \parr B$, $A \multimap B$, is called the multiplicative
fragment of linear logic, or MLL. It talks about the basics of
transactions, loading up addresses with resources and establishing
dependencies between addresses, essentially recording transaction
history. However, it does so in a way that keeps track of how the
blockchain itself is segmented. This allows us to determine things
like how much of the blockchain do i have to see in order to safely
conduct this transaction, or can i conduct this transaction without
needing visibility into that region of the blockchain.

\subsection{The Additives}

Linear logic also enjoys another fragment, called the additives. This
aspect of the logic is all about conditionals and contingencies, this
or that, but not both. The linear logic connective called 'with', and
denoted $A \with B$, collects options together into a menu for subsequent
selection by interaction with choices indicated by the linear
connective 'plus', $A + B$. In symbols,

\begin{mathpar}
  \inferrule* {\vdash \Gamma, t : A, u : B}{\vdash \Gamma, t \with u : A \with B}
\end{mathpar}

while 

\begin{mathpar}
  \inferrule* {\vdash \Gamma, t : A}{ \vdash inl( t ) : A + B }
\end{mathpar}

and

\begin{mathpar}
  \inferrule* {\vdash \Gamma, u : B}{\vdash inr( u ) : A + B}
\end{mathpar}

If during a more complex transaction $t \with u$ gets tied to $inl( t'
)$, via $txn( t \with u, inl( t' ) )$, then this will reduce to a
transaction of the form $txn( t, t' )$. On the other hand, $txn( t
\with u, inr( u' ) )$ will reduce to a transaction of the form $txn(
u, u' )$.

\subsection{The Exponentials}

The fragment of linear logic that includes the multiplicative and
additive connectives is called MALL. The remaining connectives are
called the exponentials, $?A$, and $!A$. They denote copyable,
non-conserved resources. When we write $\vdash \Gamma, t : !A$, we are
saying that you can get as many $A$'s from the address (or program)
$t$ as you want. Thus, unlike currency, that address is linked to a
copyable resource like a document, or a jpeg, or audio file, or
... that can be shared widely. When we write $\vdash \Gamma, t : ?A$, we
are saying that you can put as many $A$'s into the address (or
program) $t$ as you want. You can think of it as a place to store
$A$'s, or discard them.

What's critically important about the use of the exponentials is that
they mark resources that ought not to stay on the blockchain. They
indicate content and content types that can be better served by a
different kind of content delivery network. This is another important
function in helping with a scalable blockchain -- use blockchain
technology where it makes sense and use other means where it doesn't.

Taken all together, we have an interpretation of full classical linear
logic in terms of operations on the blockchain.

\section{Conclusions and Future Work}

We have developed a view of full classical linear logic in terms of
operations against the blockchain. The view we have been developing
not only extends to provide a meaningful interpretation of full
classical linear logic to natural and intuitive operations on the
blockchain, it also extends and expands how we think about the
blockchain and what transactions on it are. Additionally, it provides
guarantees, mathematical certainties about the correctness of
transactions structured and executed this way. In particular, notice
that we focused mostly on the connectives governing $A$'s and $B$'s (the
resources to be found at addresses or programs). We didn't really talk
about the structure of $t$'s and $u$'s. These provide us with a simple and
intuitive syntax for transactions. Of equal importance, these
transactions are \emph{typed} programs. When we write $\vdash \Gamma, t : A$, we are
not only saying something about the resources produced or manipulated
by $t$, we are saying something about how t can be used, and in what
blockchain context we can expect t to perform correctly.

Understood this way, the blockchain interpretation gives new meaning
and perspective on some theorems from the linear logic literature. In
particular, it is well established that there is a natural notion of
execution of $t$'s. That is, when thought of as programs, we know how
to run them. When they are well typed, that is, if we have established
$\vdash t : A$, then $t$ is \emph{terminating}. That's a theorem from
\cite{DBLP:journals/tcs/Abramsky93}. What this means for the
blockchain is that proof terms and their linear connectives provide a
scripting language for transactions that, on the one hand, provides
termination for all well typed scripts, and on the other is highly
expressive. Further, if it turns out that this scripting language is
not expressive enough, then there is a natural extension of proof
terms via a correspondence between linear proof terms and
{$\pi$}-calculus processes that we mentioned at the top of these
notes.

\begin{tabular}{c|c}
  proof term & blockchain meaning \\
  \hline
  address & address \\
  $t \otimes u$ & isolated concurrent transactions \\
  $t \parr u$ & interacting or linked concurrent transactions \\
  $t \with u$ & menu of transaction options \\
  $inl( t ), inr( u )$ & transaction option selection \\
  $!t$ & copyable resource server \\
  $?t$ & copyable resource storage \\
  $txn( t, u )$ & joined transactions\\
\end{tabular}

This correspondence is not just useful for extending a scripting
language for blockchain transactions. It turns out the
{$\pi$}-calculus the premier formalism for specifying, reasoning
about, and executing protocols in distributed systems
\cite{milner91polyadicpi} \cite{DBLP:journals/mscs/Milner92}
\cite{DBLP:conf/aplas/AbadiCF06} \cite{DBLP:journals/tissec/AbadiBF07}
\cite{DBLP:conf/lics/Gordon06} \cite{DBLP:conf/esop/KremerR05}. Since
one of the real values of the blockchain is the fact that it is a
distributed means to conduct transactions, the need to tie this
formalism to one for specifying protocols in distributed systems is
plain.

\subsection{Proof-of-work}

The glaring lacunae in this discussion is, of course, the relationship
to proof-of-work. Consider the following example. Suppose $C_1$ and $C_2$
are blockchains both of height $N$.

\begin{equation*}
  \begin{aligned}
    C_1 = B_{1N} \leftarrow B_{1N-1} \leftarrow \ldots \leftarrow B_{10} \\
    C_2 = B_{2N} \leftarrow B_{2N-1} \leftarrow \ldots \leftarrow B_{20}
  \end{aligned}
\end{equation*}

We can define

\begin{equation*}
  C_1 \otimes C_2 = ( B_{1N} \otimes B_{2N} ) \leftarrow ( B_{1N-1} \otimes B_{2N-1} ) \leftarrow \ldots \leftarrow ( B_{10} \otimes B_{20} )
\end{equation*}

Note that it is insufficient merely to guarantee for $B \otimes B'$ that all
the transactions in $B$ are isolated from the transactions in $B'$. The
counterexample is

\begin{equation*}
  \begin{aligned}
    C_1 = Block\{ 1AliceAddr \xrightarrow{5btc} 1AllanAddr \} \\
    \leftarrow Block\{ 1BobAddr \xrightarrow{7btc} 1BettyAddr \} \\
    C_2 = Block\{ 1BobAddr \xrightarrow{7btc} 1BettyAddr \} \\
    \leftarrow Block\{ 1AliceAddr \xrightarrow{5btc} 1AllanAddr \}
  \end{aligned}
\end{equation*}

Clearly $B_{11}$ is isolated from $B_{21}$, and $B_{10}$ is isolated
from $B_{20}$; but, $B_{20}$ is not isolated from $B_{11}$, and
$B_{10}$ is not isolated from $B_{21}$. As a result, the spends in the
earlier blocks could impact the spends in the later blocks.

Instead, the entire address space of $C_1$ must be isolated from
$C_2$. In this case the network of servers, $N_1$, that maintain $C_1$ can
be safely combined with the network of servers, $N_2$, that maintain $C_2$,
and we can safely define the composite chain as above. The
proof-of-work protocol organizing $N_1$ is completely separate from that
in $N_2$. They do not interact. Yet, it is safe to combine the chains
using a glorified zip function. In this example,

\begin{equation*}
  \begin{aligned} 
    C_1 = Block\{ 1AliceAddr \xrightarrow{5btc} 1AllanAddr \} \\
    \leftarrow Block\{ 2BobAddr \xrightarrow{7btc} 2BettyAddr \} \\
    C2 = Block\{ 1BobAddr \xrightarrow{7btc} 1BettyAddr \} \\
    \leftarrow Block\{ 2AliceAddr \xrightarrow{5btc} 2AllanAddr \}
  \end{aligned}
\end{equation*}

The address spaces of these chains are completely isolated (often
written $addresses( C_1 ) \# addresses( C_2 )$ ). We are free to calculate

\begin{equation*}
  \begin{aligned} 
    C_1 \otimes C_2  = Block\{ 1AliceAddr \xrightarrow{5btc} 1AllanAddr \} \\
    \otimes Block\{ 1BobAddr \xrightarrow{7btc} 1BettyAddr \}  \\
    \leftarrow Block\{ 2BobAddr \xrightarrow{7btc} 2BettyAddr \} \\
    \otimes Block\{ 2AliceAddr \xrightarrow{5btc} 2AllanAddr \}\\
    = Block\{ 1AliceAddr \xrightarrow{5btc} 1AllanAddr ; \\
    1BobAddr \xrightarrow{7btc} 1BettyAddr \} \\
    \leftarrow Block\{ 2BobAddr \xrightarrow{7btc} 2BettyAddr \\
    ; 2AliceAddr \xrightarrow{5btc} 2AllanAddr \}
  \end{aligned}
\end{equation*}

The ordering of transactions provided by the two independently
executing proof-of-work protocols is combined in a completely safe.

Note that there are at least two possible interpretations of $C_1 *
C_2$. One is that the requirement is to verify that $addresses( C_1 ) \#
addresses( C_2 )$. Another is to ensure this is the case by rewiring the
transactions. Under this latter interpretation even the counterexample
becomes safe

\begin{equation*}
  \begin{aligned} 
    C_1 \otimes C_2  =  Block\{ 1AliceAddr \xrightarrow{5btc} 1AllanAddr \} \\
    \otimes Block\{ 1BobAddr \xrightarrow{7btc} 1BettyAddr \} \\
    \leftarrow Block\{ 1BobAddr \xrightarrow{7btc} 01BettyAddr \} \\
    \otimes Block\{ 1AliceAddr \xrightarrow{5btc} 11AllanAddr \} \\
    = Block\{ 01AliceAddr \xrightarrow{5btc} 01AllanAddr ; \\
    11BobAddr \xrightarrow{7btc} 11BettyAddr \} \\
    \leftarrow Block\{ 01BobAddr \xrightarrow{7btc} 01BettyAddr ; \\
    11AliceAddr \xrightarrow{5btc} 11AllanAddr \}
  \end{aligned}
\end{equation*}

There is much more to be said, but that must be left to future work!

\paragraph{Acknowledgments}
We would like to acknowledge Vlad Zamfir for some thoughtful and
stimulating conversation about the blockchain protocol.

\bibliographystyle{amsplain}
\bibliography{ltcctbc}


\section{Appendix: a terminating scripting language}

In the main body of the paper we presented what amounts to the high
level intuitions. In this appendix we present enough of the details
that a reader skilled in the art could implement the proposal to test
it for themselves. This presentation follows Abramksy's proof
expressions from \cite{DBLP:journals/tcs/Abramsky93} very closely.

\subsection{Syntax}
\begin{grammar}
{p,q} \bc \mathsf{(} e_1, \ldots, e_m \mathsf{)} \mathsf{\{} t_1 \mathsf{;} \; \ldots \mathsf{;} \; t_n \mathsf{\}} & \mbox{programs} \\
{e} \bc \mathsf{satoshi} \; \bm \; \ldots \; \bm \; \mathsf{ampere} & \mbox{currency units} \\
    \;\;\; \bm \; x & \mbox{address} \\
    \;\;\; \bm \; e * e & \mbox{isolation} \\
    \;\;\; \bm \; e \# e  & \mbox{connection} \\
    \;\;\; \bm \; e \multimap e  & \mbox{obligation} \\
    \;\;\; \bm \; \mathsf{choose}\mathsf{(} x_1, \ldots, x_n \mathsf{)}\mathsf{\{} p \mathsf{;}\; q \mathsf{\}} & \mbox{menu} \\
    \;\;\; \bm \; \mathsf{inl}\mathsf{(} e \mathsf{)} \; \bm \; \mathsf{inr}\mathsf{(} e \mathsf{)} & \mbox{selection} \\
    \;\;\; \bm \; \mathsf{?}e \; \bm \; \mathsf{\_} & \mbox{storage, disposal} \\
    \;\;\; \bm \; e \mathsf{@} e & \mbox{contraction} \\
    \;\;\; \bm \; \mathsf{!}\mathsf{(} x_1, \ldots, x_n \mathsf{)}\mathsf{\{} p \mathsf{\}} & \mbox{replication} \\
{t} \bc \mathsf{txn}\mathsf{(} e_1, e_2 \mathsf{)} & \mbox{transaction}
\end{grammar}

\paragraph{Discussion}
$e_1 \multimap e_2$ is really just convenient syntactic sugar for
$e_1^{\perp} \# e_2$, where $e^{\perp}$ is identity on addresses, but
changes the \emph{polarity} of the type and otherwise operates as

\begin{equation*}
  \begin{aligned}
    ( e_1 * e_2 )^{\perp} = e_1^{\perp} \# e_2^{\perp} \\
    ( e_1 \# e_2 )^{\perp} = e_1^{\perp} * e_2^{\perp}
  \end{aligned}
\end{equation*}

\subsubsection{Interpretation}
Programs $p$ and $q$ represent blockchain states. For $p = \mathsf{(}
e_1, \ldots, e_m \mathsf{)} \mathsf{\{} t_1 \mathsf{;} \; \ldots
\mathsf{;} \; t_n \mathsf{\}}$, the $e$'s represent resources
available on the blockchain $p$, while the $t$'s represent
transactions in progress. For example, if we write $M \cdot \mathsf{satoshi}$ for $\underbrace{\mathsf{satoshi} * \ldots * \mathsf{satoshi}}_{M}$, then 

$\mathsf{(} 1blkchnaddr {)} \mathsf{\{} \mathsf{txn}\mathsf{(} 1blkchnaddr, M \cdot \mathsf{satoshi} \mathsf{)} \mathsf{\}}$

represents the genesis block where $1blkchnaddr$ has been assigned $M$
$\mathsf{satoshi}$'s. At the other end of the spectrum,

$\mathsf{(} 1blkchnaddr {)} \mathsf{\{} \mathsf{txn}\mathsf{(} 1blkchnaddr, \mathsf{\_} \mathsf{)} \mathsf{\}}$

represents burning the assets sent to $1blkchnaddr$.

At this level of abstraction modeled by the operational semantics in
the next section, addresses are more closely aligned with transaction
inputs in blockchain transactions. Thus, the genesis block is more
accurately represented as 

\begin{alignat*}{2}
  &\mathsf{(} addr_1 * \ldots * addr_M {)} \mathsf{\{} && \\
  &\;\;\;\;\mathsf{txn}\mathsf{(} addr_1, \mathsf{satoshi} \mathsf{)} \mathsf{;} && \\
  &\;\;\;\;\ldots \mathsf{;} && \\
  &\;\;\;\;\mathsf{txn}\mathsf{(}addr_M, \mathsf{satoshi} \mathsf{)} && \\
  &\mathsf{\}} &&
\end{alignat*} 

which for future reference we'll write $\mathsf{genesis}$. Similarly,
the second example is more accurately written as

\begin{equation*}
  \begin{aligned} 
    \mathsf{(} addr_1 * \ldots * addr_M {)} \mathsf{\{} \mathsf{txn}\mathsf{(} addr_1, \mathsf{\_} \mathsf{)} \mathsf{;} \ldots \mathsf{;} \; \mathsf{txn}\mathsf{(}addr_M, \mathsf{\_} \mathsf{)} \mathsf{\}}
  \end{aligned} 
\end{equation*}

we'll write as $\mathsf{burn}$ in the sequel.

\subsection{Operational Semantics}

In what follows we use the notational conventions:
\begin{itemize}
  \item $\vec{e}$ is a list of $e$'s of length $|\vec{e}|$; likewise $\vec{t}$ is list of $t$'s.
  \item $\mathsf{txn}\mathsf{(} \vec{e}, \vec{e'} \mathsf{)} = \mathsf{txn}\mathsf{(} e_1, e_1' \mathsf{)}\mathsf{;}\; \ldots \mathsf{;} \; \mathsf{txn}\mathsf{(} e_n, e_n' \mathsf{)}$ assuming $|\vec{e}| = |\vec{e'}|$
  \item we have operations, $( - )^l : \mathsf{Addr} \to \mathsf{Addr}$, $( - )^r : \mathsf{Addr} \to \mathsf{Addr}$ such that given an address $x$, $x^l$, $x^r$ are distinct from $x$ and each other; these operations extend uniquely to $p$, $e$, and $t$ in the obvious manner.
\end{itemize}

\begin{mathpar}
  \inferrule* [lab=Transaction] {}{\mathsf{txn}\mathsf{(} e_1, x \mathsf{)}\mathsf{;} \; \mathsf{txn}\mathsf{(} x, e_2 \mathsf{)} \rightarrow \mathsf{txn}\mathsf{(} e_1, e_2 \mathsf{)}}
\end{mathpar}

\begin{mathpar}
  \inferrule* [lab=Pair] {}{\mathsf{txn}\mathsf{(} e_1 * e_1', e_2 \# e_2' \mathsf{)} \rightarrow \mathsf{txn}\mathsf{(} e_1, e_2 \mathsf{)}\mathsf{;} \; \mathsf{txn}\mathsf{(} e_1', e_2' \mathsf{)}}
\end{mathpar}

\begin{mathpar}
  \inferrule* [lab=Left] {}{\mathsf{txn}\mathsf{(} \mathsf{choose}\mathsf{(} x, \vec{x} \mathsf{)}\mathsf{\{} \mathsf{(} e, \vec{e} \mathsf{)} \mathsf{\{} \vec{t} \mathsf{\}} \mathsf{;} \; q \mathsf{\}}, \mathsf{inl}\mathsf{(} e' \mathsf{)} \mathsf{)} \\ \rightarrow \mathsf{txn}\mathsf{(} e, e' \mathsf{)}\mathsf{;} \; \vec{t} \mathsf{;} \; \mathsf{txn}\mathsf{(} \vec{x}, \vec{e} \mathsf{)}}
\end{mathpar}

\begin{mathpar}
  \inferrule* [lab=Right] {}{\mathsf{txn}\mathsf{(} \mathsf{choose}\mathsf{(} x, \vec{x} \mathsf{)}\mathsf{\{} p \mathsf{;}\; \mathsf{(} e, \vec{e} \mathsf{)} \mathsf{\{} \vec{t} \mathsf{\}} \mathsf{\}}, \mathsf{inr}\mathsf{(} e' \mathsf{)} \mathsf{)} \\ \rightarrow \mathsf{txn}\mathsf{(} e, e' \mathsf{)}\mathsf{;} \; \vec{t} \mathsf{;} \; \mathsf{txn}\mathsf{(} \vec{x}, \vec{e} \mathsf{)}}
\end{mathpar}

\begin{mathpar}
  \inferrule* [lab=Read] {}{\mathsf{txn}\mathsf{(} \mathsf{!}\mathsf{(} \vec{x} \mathsf{)}\mathsf{\{} \mathsf{(} e, \vec{e} \mathsf{)} \mathsf{\{} \vec{t} \mathsf{\}} \mathsf{\}}, \mathsf{?}e' \mathsf{)} \rightarrow \mathsf{txn}\mathsf{(} e, e' \mathsf{)}\mathsf{;} \; \mathsf{txn}\mathsf{(} \vec{x}, \vec{e} \mathsf{)}}
\end{mathpar}

\begin{mathpar}
  \inferrule* [lab=Dispose] {}{\mathsf{txn}\mathsf{(} \mathsf{!}\mathsf{(} \vec{x} \mathsf{)}\mathsf{\{} p \mathsf{\}}, \mathsf{\_} \mathsf{)} \rightarrow \mathsf{txn}\mathsf{(} \vec{x}, \mathsf{\_} \mathsf{)}}
\end{mathpar}

\begin{mathpar}
  \inferrule* [lab=Copy] {}{\mathsf{txn}\mathsf{(} \mathsf{!}\mathsf{(} \vec{x} \mathsf{)}\mathsf{\{} p \mathsf{\}}, e_1 \mathsf{@} e_2 \mathsf{)} \\ \rightarrow \mathsf{txn}\mathsf{(} \vec{x}, x^l \mathsf{@} x^r \mathsf{)} \mathsf{;} \; \mathsf{txn}\mathsf{(} \mathsf{!}\mathsf{(} \vec{x} \mathsf{)}\mathsf{\{} p \mathsf{\}}^l, e_1\mathsf{)} \mathsf{;} \; \mathsf{txn}\mathsf{(} \mathsf{!}\mathsf{(} \vec{x} \mathsf{)}\mathsf{\{} p \mathsf{\}}^r, e_2\mathsf{)}}
\end{mathpar}

\subsubsection{Interpretation}

The operational semantics should be viewed as the specification of an
abstract machine that needs no other registers than the program
itself. Let's look at an example in some detail.

Executing a transaction amounts to joining to expressions, $e_1$ and
$e_2$ in $\mathsf{txn}\mathsf{(} e_1, e_2\mathsf{)}$. Thus, to send
$I < M$ $\mathsf{satoshi}$ to $bddr_1 * \ldots * bddr_I$, in the context of the genesis
block, first we have to turn the genesis block into an expression.

$\mathsf{choose}\mathsf{(} 1spndaddr \mathsf{)}\mathsf{\{} \mathsf{genesis}\mathsf{;}\; \mathsf{burn} \mathsf{\}}$

Next, we form a spend expression $bddr_1 * \ldots * bddr_I \multimap addr_{I+1} \# addr_M$ which will consume $I$ $\mathsf{satoshi}$ from the genesis block addresses $addr_1$ through $addr_I$, and deposit them in $addr_1$ through $addr_I$.

Now, we can create a transaction that selects the genesis block from
the menu of blockchain states via

\begin{alignat*}{2}
  &\mathsf{txn}\mathsf{(} && \\
  &\;\;\;\;\mathsf{choose}\mathsf{(} 1spndaddr \mathsf{)}\mathsf{\{} \mathsf{genesis}\mathsf{;}\; \mathsf{burn} \mathsf{\}}, && \\
  &\;\;\;\;\mathsf{inl}\mathsf{(} bddr_1 * \ldots * bddr_I \multimap addr_{I+1} \# addr_M \mathsf{)} && \\
  &\mathsf{)}
\end{alignat*}

Using the operational semantics we see that this reduces to

\begin{alignat*}{2}
  & \mathsf{txn}\mathsf{(} && \\
  & \;\;\;\; addr_1 * \ldots * addr_M, && \\
  & \;\;\;\; bddr_1 * \ldots * bddr_I \multimap addr_{I+1} \# addr_M && \\
  & \mathsf{)} \mathsf{;}
\end{alignat*}

which then reduces to

\begin{alignat*}{2}
  &\mathsf{txn}\mathsf{(} addr_1, bddr_1 \mathsf{)} \mathsf{;} && \\
  &\ldots \mathsf{;} && \\
  &\mathsf{txn}\mathsf{(} addr_I, bddr_I \mathsf{)} \mathsf{;} && \\
  &\mathsf{txn}\mathsf{(} addr_{I+1}, addr_{I+1} \mathsf{)} \mathsf{;} && \\
  &\ldots \mathsf{;} && \\
  &\mathsf{txn}\mathsf{(} addr_M, addr_M \mathsf{)} \mathsf{;} && \\
  &\mathsf{txn}\mathsf{(} addr_1, \mathsf{satoshi} \mathsf{)} \mathsf{;} && \\
  &\ldots \mathsf{;} && \\
  &\mathsf{txn}\mathsf{(} addr_M, \mathsf{satoshi} \mathsf{)} \mathsf{;} &&
\end{alignat*}

which then reduces to

\begin{alignat*}{2}
  &\mathsf{txn}\mathsf{(} bddr_1, \mathsf{satoshi} \mathsf{)} \mathsf{;} && \\
  &\ldots \mathsf{;} && \\
  &\mathsf{txn}\mathsf{(} bddr_I, \mathsf{satoshi} \mathsf{)} \mathsf{;} && \\
  &\mathsf{txn}\mathsf{(} addr_{I+1}, \mathsf{satoshi} \mathsf{)} \mathsf{;} && \\
  &\ldots \mathsf{;} && \\
  &\mathsf{txn}\mathsf{(} addr_M, \mathsf{satoshi} \mathsf{)} \mathsf{;} &&
\end{alignat*} 

This can be seen as a ledger-like representation assigning
$\mathsf{satoshi}$'s to addresses.

Now, the final piece of the puzzle is that that spend transaction
needs to be created in the context of a blockchain state, which
constitutes the \emph{resulting} blockchain state. In point of fact,
this is a piece of context we elided when we formed the transaction to
focus on the reduction. A more complete picture of the execution looks like

\begin{alignat*}{2}
  &\mathsf{(} bddr_1 * \ldots * bddr_I * addr_{I+1} * \ldots * addr_M  \mathsf{)} \mathsf{\{} && \\
  &\;\;\;\;\mathsf{txn}\mathsf{(} && \\
  &\;\;\;\;\;\;\mathsf{choose}\mathsf{(} 1spndaddr \mathsf{)}\mathsf{\{} \mathsf{genesis}\mathsf{;}\; \mathsf{burn} \mathsf{\}}, && \\
  &\;\;\;\;\;\;\mathsf{inl}\mathsf{(} bddr_1 * \ldots * bddr_I \multimap addr_{I+1} \# addr_M \mathsf{)} && \\
  &\;\;\mathsf{)} && \\
  &\mathsf{\}} && \\
  &\rightarrow* && \\
  &\mathsf{(} bddr_1 * \ldots * bddr_I * addr_{I+1} * \ldots * addr_M  \mathsf{)} \mathsf{\{} && \\
  &\;\;\;\;\mathsf{txn}\mathsf{(} bddr_1, \mathsf{satoshi} \mathsf{)} \mathsf{;}&& \\
  &\;\;\;\;\ldots \mathsf{;} && \\
  &\;\;\;\;\mathsf{txn}\mathsf{(} bddr_I, \mathsf{satoshi} \mathsf{)} \mathsf{;} && \\
  &\;\;\;\;\mathsf{txn}\mathsf{(} addr_{I+1}, \mathsf{satoshi} \mathsf{)} \mathsf{;} && \\
  &\;\;\;\;\ldots \mathsf{;} && \\
  &\;\;\;\;\mathsf{txn}\mathsf{(} addr_M, \mathsf{satoshi} \mathsf{)} \mathsf{;} && \\
  &\mathsf{\}} &&
\end{alignat*}

This brings us full circle. At the beginning of the paper we
explicitly recognized the blockchain as data that is program. The
reduction above provides an explicit model of just this phenomenon. A
blockchain state, i.e. a representation of data, is a
\emph{program}. The transition from one state to the next is the
execution of the program. Any state of the program actually allows a
``read back'' to a ledger-like representation capturing the
distribution of resources to addresses.

\subsection{Type assignment}

In the main body of the paper we wrote proof rules in terms of
sequents. In point of fact, that formalism amounts to a typing
discipline on the scripting language presented above. Here we present
the details of that typing discipline along with the basic result that
all well typed programs are terminating.

\begin{mathpar}
  \inferrule* [lab=Axiom] {}{\vdash \mathsf{(} x : A^{\perp}, x : A \mathsf{)} \mathsf{\{} \mathsf{\}}}
\end{mathpar}

\begin{mathpar}
  \inferrule* [lab=Tensor] {\vdash \mathsf{(} t : A, \Gamma \mathsf{)} \mathsf{\{} \vec{txn} \mathsf{\}} \\ \vdash \mathsf{(} u : B, \Delta \mathsf{)} \mathsf{\{} \vec{txn'} \mathsf{\}}}{\vdash \mathsf{(} t * u : A \otimes B, \Gamma, \Delta \mathsf{)} \mathsf{\{} \vec{txn}\mathsf{;} \; \vec{txn'} \mathsf{\}}}
\end{mathpar}

\begin{mathpar}
  \inferrule* [lab=Par] {\vdash \mathsf{(} t : A, u : B, \Gamma \mathsf{)} \mathsf{\{} \vec{txn} \mathsf{\}}}{ \vdash \mathsf{(} t \# u : A \parr B, \Gamma \mathsf{)} \mathsf{\{} \vec{txn} \mathsf{\}}}
\end{mathpar}

\begin{mathpar}
  \inferrule* [lab=With] {\vdash p \and \vdash q \\ p = \mathsf{(} t : A, \vec{t} : \vec{G} \mathsf{)} \mathsf{\{} \vec{txn} \mathsf{\}} \and q = \mathsf{(} u : B, \vec{u} : \vec{G} \mathsf{)} \mathsf{\{} \vec{txn'} \mathsf{\}}}{\vdash \mathsf{(} \mathsf{choose}\mathsf{(} \vec{x} : \vec{G} \mathsf{)}\mathsf{\{} p \mathsf{\}} \mathsf{\{} q \mathsf{\}} \mathsf{)} : A \with B, \vec{x} : \vec{G} \mathsf{)} \mathsf{\{} \mathsf{\}}}
\end{mathpar}

\begin{mathpar}
  \inferrule* [lab=Left] {\vdash \mathsf{(} t : A, \Gamma \mathsf{)} \mathsf{\{} \vec{txn} \mathsf{\}}}{ \vdash \mathsf{(} inl( t ) : A + B, \Gamma \mathsf{)} \mathsf{\{} \vec{txn} \mathsf{\}} }
  \and
  \inferrule* [lab=Right] {\vdash \mathsf{(} u : B, \Gamma \mathsf{)} \mathsf{\{} \vec{txn} \mathsf{\}}}{ \vdash \mathsf{(} inr( u ) : A + B, \Gamma \mathsf{)} \mathsf{\{} \vec{txn} \mathsf{\}} }
\end{mathpar}

\begin{mathpar}
  \inferrule* [lab=Storage] {\vdash \mathsf{(} t : A, \Gamma \mathsf{)} \mathsf{\{} \vec{txn} \mathsf{\}}}{ \vdash \mathsf{(} ?t : ?A, \Gamma \mathsf{)} \mathsf{\{} \vec{txn} \mathsf{\}} }
  \and
  \inferrule* [lab=Disposal] {\vdash \mathsf{(} t : A, \Gamma \mathsf{)} \mathsf{\{} \vec{txn} \mathsf{\}}}{ \vdash \mathsf{(} \_ : ?A, \Gamma \mathsf{)} \mathsf{\{} \vec{txn} \mathsf{\}} }
\end{mathpar}

\begin{mathpar}
  \inferrule* [lab=Contraction] {\vdash \mathsf{(} t : ?A, u : ?A, \Gamma \mathsf{)} \mathsf{\{} \vec{txn} \mathsf{\}}}{ \vdash \mathsf{(} t @ u : ?A, \Gamma \mathsf{)} \mathsf{\{} \vec{txn} \mathsf{\}}}
\end{mathpar}

\begin{mathpar}
  \inferrule* [lab=Replication] {\vdash p \\\\ p = \mathsf{(} t : A, \vec{t} : ?\vec{G}, \Gamma \mathsf{)} \mathsf{\{} \vec{txn} \mathsf{\}}}{ \vdash \mathsf{(} \mathsf{!} \mathsf{(} \vec{x} \mathsf{)} \mathsf{\{} p \mathsf{\}} : !A, \vec{x} : ?\vec{G} \mathsf{)} \mathsf{\{} \mathsf{\}}}
\end{mathpar}

\paragraph{Discussion}
As is easily seen, this is merely a transliteration of Abramsky's
proof expressions from \cite{DBLP:journals/tcs/Abramsky93}, and as
such the scripting language enjoys all the properties of proof
expressions. In particular, theorem 7.18 pg 47 tells us that well
typed programs terminate.

In a discussion of ``smart contract'' the types play a specially
important role. If programs in this language consitute financial
contracts, then the types provide a means by which parties can probe
the contracts for properties above and beyond termination.


\end{document}